# Probabilistic Analysis and Empirical Validation of Patricia Tries in Ethereum State Management

Oleksandr Kuznetsov [1,2,3*]

[1] Zpoken, Harju maakond, Tallinn, Kesklinna linnaosa, Sakala tn 7-2, 10141, Estonia

[2] Department of Theoretical and Applied Sciences, eCampus University, Via Isimbardi 10, Novedrate (CO), 22060, Italy

[3] Department of Information and Communication Systems Security, V. N. Karazin Kharkiv National University, 4 Svobody Sq., 61022 Kharkiv, Ukraine

Email: oleksandr.k@zpoken.io, oleksandr.kuznetsov@uniecampus.it, kuznetsov@karazin.ua
https://orcid.org/0000-0003-2331-6326

Anton Yezhov [1]

[1] Zpoken, Harju maakond, Tallinn, Kesklinna linnaosa, Sakala tn 7-2, 10141, Estonia

Email: anton.yezhov@zpoken.io
https://orcid.org/0009-0004-6380-5233

Kateryna Kuznetsova [1, 4]

[1] Zpoken, Harju maakond, Tallinn, Kesklinna linnaosa, Sakala tn 7-2, 10141, Estonia

[4] VRAI - Vision, Robotics and Artificial Intelligence Lab, Via Brecce Bianche 12, 60131, Ancona, Italy

Email: kateryna.k@zpoken.io, kate7smith12@gmail.com
https://orcid.org/0000-0002-5605-9293

Oleksandr Domin [1]

[1] Zpoken, Harju maakond, Tallinn, Kesklinna linnaosa, Sakala tn 7-2, 10141, Estonia

Email: scmaster@zpoken.io
https://orcid.org/0009-0009-1591-137X

*Corresponding author. E-mail(s): oleksandr.k@zpoken.io, oleksandr.kuznetsov@uniecampus.it, kuznetsov@karazin.ua

**Abstract**

This study presents a comprehensive theoretical and empirical analysis of Patricia tries, the fundamental data structure underlying Ethereum's state management system. We develop a probabilistic model characterizing the distribution of path lengths in Patricia tries containing random Ethereum addresses and validate this model through extensive computational experiments.

Our findings reveal the logarithmic scaling of average path lengths with respect to the number of addresses, confirming a crucial property for Ethereum's scalability. The study demonstrates high precision in predicting average path lengths, with discrepancies between theoretical and experimental results not exceeding 0.01 across tested scales from 100 to 100,000 addresses. We identify and verify the right-skewed nature of path length distributions, providing insights into worst-case scenarios and informing optimization strategies. Statistical analysis, including chi-square goodness-of-fit tests, strongly supports the model's accuracy. The research offers structural insights into node concentration at specific trie levels, suggesting avenues for optimizing storage and retrieval mechanisms. These findings contribute to a deeper understanding of Ethereum's fundamental data structures and provide a solid foundation for future optimizations. The study concludes by outlining potential directions for future research, including investigations into extreme-scale behavior, dynamic trie performance, and the applicability of the model to non-uniform address distributions and other blockchain systems.

## 1. Introduction

### 1.1 Patricia Tries in Ethereum: A Critical Data Structure

The Ethereum blockchain, as one of the foremost platforms for decentralized applications and smart contracts, relies heavily on efficient data structures for state management [1], [2]. At the core of Ethereum's state storage system lies the Patricia trie, a specialized variant of the radix trie optimized for efficient storage and retrieval of key-value pairs [3]. In the context of Ethereum, Patricia tries, often referred to as Modified Merkle Patricia Tries (MPT), serve a crucial role in maintaining the global state, transaction receipts, and storage of smart contracts [4].

Patricia tries in Ethereum offer several advantages [5], [6]:

1. Deterministic Root Hash: Any change in the trie results in a new root hash, providing a concise and verifiable representation of the entire state.

2. Efficient Proofs: The structure allows for the generation of compact Merkle proofs, facilitating light client implementations and state verification.

3. Incremental Updates: Patricia tries support efficient incremental updates, crucial for Ethereum's frequent state changes.

However, the performance characteristics of Patricia tries, particularly in terms of path lengths, have significant implications for Ethereum's scalability and efficiency.

### 1.2 The Critical Nature of Trie Structure Analysis

Understanding the structural properties of Patricia tries is paramount for several reasons [3], [4]:

1. Scalability Concerns: As the Ethereum network grows, the efficiency of state storage and retrieval becomes increasingly critical. The structure of Patricia tries directly impacts the performance of these operations.

2. Optimization Opportunities: In-depth knowledge of trie structures can inform optimizations in node storage, caching strategies, and state pruning mechanisms.

3. Client Implementation: Both full nodes and light clients rely on efficient trie traversal. The distribution of path lengths affects proof sizes and verification times.

4. Protocol Upgrades: Proposals for Ethereum upgrades, such as state expiry and stateless clients, necessitate a thorough understanding of the current state structure [7].

5. Resource Estimation: Accurate models of trie structures enable better estimation of computational and storage resources required for network participation.

**1.3 Addressing a Critical Research Gap**

Despite the fundamental importance of Patricia tries in Ethereum, there is a notable lack of comprehensive studies focusing on the distribution of path lengths and the overall structure of these tries. This research gap is particularly significant given the direct impact of path lengths on the efficiency of Ethereum's core operations.

Vitalik Buterin, co-founder of Ethereum, highlighted this issue in a recent discussion on potential protocol upgrades [7]:

- "The goal is to replace the current Merkle Patricia state tree (MPT), because the current MPT is *very* unfriendly to stateless clients: a worst-case stateless proof for an Ethereum block is ~300 MB (think: spam reads on 24kB contracts), and even the average case sucks because the tree is width-16."

This statement underscores the critical need for a deeper understanding of Patricia Trie structures, particularly in the context of emerging paradigms like stateless clients.

Our study aims to address this significant research gap by providing:

1. A rigorous probabilistic model for path length distributions in Patricia tries containing Ethereum addresses.

2. Empirical validation of the model through extensive computational experiments.

3. Insights into the structural properties of Patricia tries at various scales.

4. Analysis of the implications of trie structures on Ethereum's performance and scalability.

By filling this crucial gap in the current literature, our research contributes to the ongoing efforts to optimize Ethereum's architecture and informs future protocol upgrades. The insights derived from this study have far-reaching implications for the design of efficient state management systems in Ethereum and potentially other blockchain platforms.

The following sections will delve into the theoretical foundations of Patricia tries, present our probabilistic model, describe our experimental methodology, and discuss the implications of our findings for the future of Ethereum and blockchain technology at large.

**2. Literature Review: Patricia Tries in Blockchain and Path Length Analysis**

This section provides a focused review of the literature relevant to our study on path length distributions in Patricia tries, particularly in the context of Ethereum's state management. We highlight key contributions in the field and identify the research gaps that our work addresses.

## 2.1 Theoretical Foundations of Patricia Tries

The study of Patricia tries has a rich history in computer science, with several seminal works providing the theoretical groundwork for our research:

- Kirschenhofer et al. [8] conducted a pioneering analysis of the balance property in Patricia tries, focusing on external path length. Their work demonstrated that for binary Patricia trees, the variance of the external path length is asymptotically equal to $0.37...n + nP(\log_2 n)$, where n is the number of stored records. This result established that the external path length is asymptotically equal to $n \cdot \log_2 n$ with high probability, providing crucial insights into the structure's efficiency.
- Building on this, Andersson [9] offered additional perspectives on the balance property of Patricia tries, further refining our understanding of their structural characteristics. These early studies laid the foundation for understanding the efficiency and balance properties of Patricia tries, which are crucial in their application to blockchain technologies.
- More recently, Tong et al. [10] introduced a smoothed analysis model for tries and Patricia tries. Their research showed that under certain perturbation conditions, the smoothed heights of both tries and Patricia tries are in $\Theta(\log n)$. This work provided a more nuanced understanding of these structures' performance in practical scenarios, bridging the gap between worst-case and average-case analyses.

## 2.2 Patricia Tries in Blockchain and Ethereum

The application of Patricia tries in blockchain technology, particularly in Ethereum, has been a subject of intense research in recent years:

- Tabatabaei et al. [11] provided a comprehensive comparison of blockchain systems, including Ethereum, which uses a Modified Merkle Patricia Trie (MPT) for state storage. Their work offered valuable insights into how Patricia tries are adapted for blockchain use, highlighting the unique characteristics of Ethereum's storage layer.
- Mardiansyah et al. [12] proposed the Multi-State Merkle Patricia Trie (MSMPT) for multi-query processing in lightweight blockchains. Their experimental results demonstrated significant improvements in query processing performance and storage efficiency, showcasing the potential for further optimizations of Patricia trie structures in blockchain contexts.
- Yang et al. [13] addressed performance bottlenecks in Ethereum's storage engine by proposing SolsDB, a single-level ordered log structure database. Their approach demonstrated significant improvements in read performance and write amplification, highlighting ongoing efforts to optimize storage structures in blockchain systems.

## 2.3 Data Management and Security in Blockchain Systems

Several studies have explored broader aspects of data management and security in blockchain systems, providing context for our work:

- Kostamis et al. [14] conducted a thorough examination of various data management methods in Ethereum, including on-chain solutions and hybrid architectures. Their work evaluated these methods in terms of storage cost and retrieval latency, providing valuable insights into the trade-offs involved in different data management approaches.
- Liang et al. [15] proposed the BI-TSFID framework, combining Ethereum and IPFS technologies to optimize data storage and verification in IoT scenarios. Their work included refinements to the Merkle Tree structure, demonstrating the ongoing evolution of tree-based data structures in blockchain applications.

**2.4 Research Gaps and Our Contributions**

While the existing literature provides valuable insights into Patricia tries and their applications in blockchain, several key gaps remain:

- Lack of Comprehensive Path Length Analysis: Despite the fundamental importance of path lengths in Patricia tries for Ethereum's performance, there is a notable absence of studies focusing specifically on the distribution of path lengths in the context of Ethereum addresses.

- Limited Empirical Validation: Most theoretical models of Patricia tries lack extensive empirical validation, especially in the context of large-scale blockchain systems like Ethereum.

- Insufficient Focus on Ethereum-Specific Optimizations: While general optimizations for Patricia tries have been proposed, there is a lack of research tailored specifically to the unique characteristics of Ethereum's state trie.

- Absence of Predictive Models: There is a lack of models that can accurately predict the structural properties of Patricia tries as the Ethereum state grows, which is crucial for long-term scalability planning.

Our study addresses these gaps by:

- Providing a comprehensive probabilistic model for path length distributions in Patricia tries containing Ethereum addresses.

- Conducting extensive empirical validation of our model using large-scale simulations of Ethereum-like address spaces.

- Analyzing the implications of our findings specifically for Ethereum's state management and proposing targeted optimizations.

- Developing predictive models that can estimate the structural properties of Patricia tries as the Ethereum state expands.

By addressing these gaps, our research contributes significantly to the understanding of Patricia tries in the context of Ethereum and provides valuable insights for future optimizations and scalability improvements in blockchain state management systems.

**3. Background: Fundamentals of Patricia Tries**

Patricia tries, also known as radix tries or compact prefix trees, are an efficient data structure for storing and retrieving string-like keys [16], [17]. In the context of Ethereum, they play a crucial role in managing state storage and transaction processing [18], [19]. This section elucidates the structure of Patricia tries and their specific application in representing Ethereum addresses.

**3.1 Structure of Patricia Tries**

Patricia tries are an optimization of standard trie structures, designed to minimize storage requirements and maximize retrieval efficiency. Unlike standard tries, which may contain nodes with only one child, Patricia tries compress these single-child nodes, resulting in a more compact structure.

### 3.1.1 Node Types

A Patricia trie typically consists of four types of nodes [3], [18]:

1. Null Node: Represented by an empty string.
2. Branch Node: Contains 17 elements: 16 for each possible hexadecimal character (0-9 and a-f) and one for a value if the node is a terminator.
3. Leaf Node: Stores a key-value pair where the key is encoded in the path.
4. Extension Node: Contains a shared nibble path and a link to another node.

### 3.1.2 Trie Operations

The primary operations on a Patricia trie are:

1. Insertion: $O(\log N)$ time complexity, where $N$ is the number of keys.
2. Lookup: Also $O(\log N)$ time complexity.
3. Deletion: $O(\log N)$ time complexity.

These operations involve traversing the trie, potentially creating new nodes, and in the case of deletion, potentially removing or merging nodes.

## 3.2 Representation of Ethereum Addresses in Patricia Tries

Ethereum leverages a modified version of Patricia tries, known as Modified Merkle Patricia Tries (MPT), to efficiently store and manage its state.

### 3.2.1 Ethereum Address Format

Ethereum addresses are 20-byte (160-bit) identifiers, typically represented as 40-character hexadecimal strings. For example:

```
0x742d35Cc6634C0532925a3b844Bc454e4438f44e
```

### 3.2.2 Trie Representation

In the Ethereum state trie, addresses serve as keys. The process of storing an address in the trie involves:

1. Key Transformation: The 20-byte address is first converted to its 40-character hexadecimal representation.
2. Path Construction: Each character of the hexadecimal string becomes a nibble in the trie path. Thus, an Ethereum address corresponds to a path of 40 nibbles in the trie.
3. Node Creation: Depending on the existing structure of the trie, the insertion of an address may result in the creation of leaf, extension, or branch nodes.

### 3.2.3 Merkle Proofs

The use of MPTs in Ethereum allows for efficient generation of Merkle proofs. Given an address, one can prove its inclusion (or non-inclusion) in the state by providing:

1. The value associated with the address (if it exists)
2. The set of nodes along the path from the root to the leaf containing the address

The Merkle proof has a space complexity of $O(\log N)$, where $N$ is the number of addresses in the trie.

### 3.2.4 Optimization Considerations

The structure of Ethereum addresses in Patricia tries has significant implications for storage and retrieval efficiency:

1. Common Prefixes: Addresses with common prefixes share paths in the trie, reducing storage requirements.
2. Balanced Structure: The pseudo-random nature of Ethereum addresses tends to create a relatively balanced trie structure, ensuring consistent lookup times.
3. Incremental Updates: The trie structure allows for efficient incremental updates, crucial for Ethereum's frequent state changes.
4. Deterministic Root Hash: Any change in the trie results in a new root hash, providing a concise representation of the entire state.

Understanding these fundamentals is crucial for analyzing the expected path lengths in Patricia tries for randomly generated Ethereum addresses, which forms the basis for subsequent sections of this study.

## 4. Probabilistic Model for Random Addresses

In order to analyze the expected structure of Patricia tries containing Ethereum addresses, it is crucial to develop a robust probabilistic model for the generation and composition of these addresses. This section elucidates the process of random address generation and examines the probabilistic distribution of characters within these addresses, providing a foundation for subsequent analyses of trie structures.

### 4.1 Description of the Random Address Generation Process

The generation of random Ethereum addresses is a process that aims to produce unique, unpredictable, and uniformly distributed identifiers within the address space [3]. This process is fundamental to the security and scalability of the Ethereum network.

#### 4.1.1 Address Space

Ethereum addresses are 20-byte (160-bit) identifiers, typically represented as 40-character hexadecimal strings [3]. The total address space is thus $2^{160}$, or approximately $1.46 \times 10^{48}$ possible addresses.

#### 4.1.2 Cryptographic Key Generation

The process of generating a random Ethereum address typically involves the following steps:

1. Private Key Generation: A 256-bit (32-byte) private key is generated using a cryptographically secure random number generator (CSPRNG). This can be represented as: $PrivateKey = CSPRNG(256)$ where $CSPRNG(n)$ denotes the output of a CSPRNG generating $n$ bits.

2. Public Key Derivation: The corresponding 512-bit public key is derived from the private key using elliptic curve cryptography, specifically the secp256k1 curve. This process can be represented as: $PublicKey = secp256k1(PrivateKey)$

3. Address Computation: The Ethereum address is computed by taking the Keccak-256 hash of the public key and retaining the last 20 bytes (160 bits). This can be expressed as: $Address = rightmost_{20}(Keccak256(PublicKey))$ where $rightmost_n(x)$ denotes the rightmost $n$ bytes of $x$.

### 4.1.3 Uniform Distribution Property

The use of cryptographic hash functions in the address generation process ensures that the resulting addresses are uniformly distributed across the address space, assuming the underlying CSPRNG is truly random. This property is crucial for our subsequent probabilistic analysis.

### 4.1.4 Collision Resistance

Given the vast address space and the uniform distribution of addresses, the probability of collision (two distinct private keys generating the same address) is negligibly small. The probability of a collision in a set of $N$ randomly generated addresses can be approximated using the birthday problem formula:

$$P(\text{collision}) \approx 1 - e^{-\frac{N^2}{2 \times 2^{160}}}.$$

where:

- $N$ is the number of randomly generated addresses,
- $2^{160}$ is the size of the possible address space.

For practical values of $N$, this probability is extremely low, allowing us to treat each generated address as unique in our model.

## 4.2 Probabilistic Distribution of Characters in Addresses

Understanding the distribution of characters within Ethereum addresses is crucial for analyzing the structure of Patricia tries containing these addresses.

### 4.2.1 Character Set

Ethereum addresses are represented as hexadecimal strings, using the character set $0, 1, 2, ..., 9, a, b, c, d, e, f$. Each character represents 4 bits of the underlying 160-bit address.

### 4.2.2 Uniform Distribution of Characters

Due to the uniform distribution property of the address generation process, each character in an Ethereum address can be modeled as an independent, identically distributed (i.i.d.) random variable. The probability of any specific hexadecimal character appearing at any position in the address is:

$$P(char = x) = \frac{1}{16}, \text{ for } x \in 0, 1, ..., f.$$

### 4.2.3 Probability of Specific Sequences

The probability of observing a specific sequence of characters in an address can be computed using the multiplication rule of probability. For a sequence of length $k$, the probability is:

$$P(sequence) = \left(\frac{1}{16}\right)^k.$$

For example, the probability of observing the sequence "a7b" at any specific position in the address is

$$\left(\frac{1}{16}\right)^3 = \frac{1}{4096}.$$

### 4.2.4 Prefix Probabilities

In the context of Patricia tries, the probability of addresses sharing a common prefix is of particular interest. The probability that two randomly chosen addresses share a prefix of length $k$ is:

$$P(shared\ prefix_k) = \left(\frac{1}{16}\right)^k.$$

This geometric distribution of shared prefixes has significant implications for the expected structure of Patricia tries containing these addresses.

Understanding these probabilistic properties of random Ethereum addresses provides a solid foundation for analyzing the expected structure and performance characteristics of Patricia tries used in Ethereum's state management system. The uniform distribution and independence of characters in these addresses lead to predictable statistical properties that can be leveraged for optimizing storage and retrieval operations in large-scale blockchain systems.

## 5. Analysis of Path Lengths

The analysis of path lengths in Patricia tries is crucial for understanding the efficiency of storage and retrieval operations in Ethereum's state management system. This section provides a rigorous examination of path lengths, beginning with a precise definition of the concept within the context of Patricia tries. Subsequent subsections will build upon this foundation to derive probabilistic models and practical implications for system performance.

### 5.1 Definition of "Path Length" in the Context of Patricia Tries

In the realm of Patricia tries, particularly those used in Ethereum's state storage, the concept of path length carries specific connotations that warrant careful definition and analysis.

#### 5.1.1 Formal Definition

Let $T$ be a Patricia trie and $k$ be a key (in our case, an Ethereum address) stored in $T$. The path length for key $k$, denoted as $PL(k,T)$, is defined as the number of nodes traversed from the root of $T$ to the leaf node containing $k$, inclusive of both the root and leaf nodes.

Mathematically, we can express this as:

$$PL(k,T) = |\{n_i : n_i \in Path(root_T, leaf_k)\}|,$$

where $Path(root_T, leaf_k)$ is the set of nodes in the path from the root of $T$ to the leaf node containing $k$, and $|\cdot|$ denotes the cardinality of the set.

### 5.1.2 Characteristics of Path Length

Several key characteristics of path length in Patricia tries merit consideration:

- Minimum Path Length: The minimum possible path length is 2, occurring when a key diverges from all others at the root node: $PL_{min}(T) = 2$.

- Maximum Path Length: The maximum path length is bounded by the number of nibbles in the key plus one (for the root). For Ethereum addresses, this is: $PL_{max}(T) \leq 41$.

- Relation to Key Prefixes: The path length for a key $k$ is directly related to the length of the longest common prefix it shares with any other key in the trie. Let $LCP(k,T)$ be the length of the longest common prefix (in nibbles) of $k$ with any other key in $T$. Then: $PL(k,T) \leq LCP(k,T) + 2$. The equality holds when each nibble of the common prefix is represented by a separate node in the trie.

### 5.1.3 Path Length and Trie Operations

The concept of path length is intimately connected to the performance of key operations in Patricia tries:

- Lookup Operation: The time complexity of a lookup operation for a key $k$ is $O(PL(k,T))$.

- Insertion Operation: Inserting a new key $k'$ into $T$ has a time complexity of $O(PL(k',T'))$, where $T'$ is the trie after insertion.

- Deletion Operation: Deleting a key $k$ from $T$ has a time complexity of $O(PL(k,T))$ plus potential rebalancing operations.

Understanding the concept of path length and its implications is fundamental to analyzing the performance characteristics of Patricia tries in Ethereum's state management system. It provides a basis for optimizing storage structures, estimating operational costs, and designing efficient scaling solutions for blockchain systems.

## 5.2 Derivation of the Formula for the Probability of a Specific Path Length

Building upon the foundational concepts established in section 4.1, we now proceed to derive a rigorous formula for the probability of a specific path length in a Patricia trie containing randomly generated Ethereum addresses.

### 5.2.1 Probability Model

Let $T$ be a Patricia trie containing $N$ randomly generated Ethereum addresses. We aim to determine $P(PL(k,T) = k)$, the probability that a randomly chosen key $k$ in $T$ has a path length of exactly $k$.

### 5.2.2 Event Decomposition

The event of a key having a path length of $k$ can be decomposed into two sub-events:

1. Event A: The first $k-1$ symbols of the key match with at least one other key in the trie.

2. Event B: The $k$-th symbol of the key does not match with any other key that shared the first $k-1$ symbols.

Thus, we can express our target probability as:

$$P(PL(k,T)=k) = P(A \cap B).$$

### 5.2.3 Complementary Events

To calculate this probability, we utilize the concept of complementary events:

1. Let $E_{k-1}$ be the event that the first $k-1$ symbols match with at least one other key.
2. Let $E_k$ be the event that the first $k$ symbols match with at least one other key.

We can then express our target probability as the difference between these two events:

$$P(PL(k,T)=k) = P(E_{k-1}) - P(E_k).$$

### 5.2.4 Probability Calculations

To calculate $P(E_k)$, we first consider its complement - the probability that the first $k$ symbols do not match with any other key:

$$P(\text{no match in first } k \text{ symbols}) = \left(1 - \left(\frac{1}{16}\right)^k \cdot \frac{15}{16}\right)^N.$$

Therefore,

$$P(E_k) = 1 - \left(1 - \left(\frac{1}{16}\right)^k \cdot \frac{15}{16}\right)^N.$$

### 5.2.5 Final Formula

Combining these elements, we arrive at our final formula:

$$P(PL(k,T)=k) = \left(1 - \left(1 - \left(\frac{1}{16}\right)^{k-1} \cdot \frac{15}{16}\right)^N\right) - \left(1 - \left(1 - \left(\frac{1}{16}\right)^k \cdot \frac{15}{16}\right)^N\right).$$

This can be simplified to:

$$P(PL(k,T)=k) = \left(1 - \left(\frac{1}{16}\right)^k \cdot \frac{15}{16}\right)^N - \left(1 - \left(\frac{1}{16}\right)^{k-1} \cdot \frac{15}{16}\right)^N.$$

This formula provides a precise probability for a key in a Patricia trie of $N$ random Ethereum addresses to have a path length of exactly $k$.

### 5.3 Calculation of Probabilities for Various Path Lengths

With our derived formula, we can now calculate and analyze the probabilities of various path lengths in Patricia tries of different sizes.

### 5.3.1 Probability Distribution

For a Patricia trie $T$ with $N$ random Ethereum addresses, the probability distribution of path lengths can be calculated as follows:

$$P(PL(k,T) = k) = \left(1 - \left(\frac{1}{16}\right)^k \cdot \frac{15}{16}\right)^N - \left(1 - \left(\frac{1}{16}\right)^{k-1} \cdot \frac{15}{16}\right)^N,$$

for $2 \leq k \leq 41$, where 41 is the maximum possible path length for an Ethereum address (40 nibbles plus the root node).

### 5.3.2 Numerical Analysis

Let's consider Patricia tries of various sizes and calculate the probabilities of different path lengths:

1. Small Trie ($N = 100$):

    - $P(PL(k,T) = 1) \approx 0.002386$;
    - $P(PL(k,T) = 2) \approx 0.690504$;
    - $P(PL(k,T) = 3) \approx 0.284479$;
    - $P(PL(k,T) = 4) \approx 0.021201$;
    - $P(PL(k,T) = 5) \approx 0.001340$;
    - $P(PL(k,T) = 6) \approx 0.000084$;
    - $P(PL(k,T) = 7) \approx 0.000005$;

2. Medium Trie ($N = 10,000$):

    - $P(PL(k,T) = 3) \approx 0.101360$;
    - $P(PL(k,T) = 4) \approx 0.765349$;
    - $P(PL(k,T) = 5) \approx 0.124390$;
    - $P(PL(k,T) = 6) \approx 0.008342$;
    - $P(PL(k,T) = 7) \approx 0.000524$;
    - $P(PL(k,T) = 8) \approx 0.000033$;
    - $P(PL(k,T) = 9) \approx 0.000002$;

3. Large Trie ($N = 1,000,000$):

    - $P(PL(k,T) = 4) \approx 0.000001$;
    - $P(PL(k,T) = 5) \approx 0.408987$;
    - $P(PL(k,T) = 6) \approx 0.536665$;
    - $P(PL(k,T) = 7) \approx 0.050860$;
    - $P(PL(k,T) = 8) \approx 0.003268$;
    - $P(PL(k,T) = 9) \approx 0.000205$;
    - $P(PL(k,T) = 10) \approx 0.000013$;
    - $P(PL(k,T) = 11) \approx 0.000001$;

4. Ethereum Tree ($N = 300,000,000$):

- $P(PL(k,T) = 7) \approx 0.350730$;
- $P(PL(k,T) = 8) \approx 0.585884$;
- $P(PL(k,T) = 9) \approx 0.059301$;
- $P(PL(k,T) = 10) \approx 0.003829$;
- $P(PL(k,T) = 11) \approx 0.000240$;
- $P(PL(k,T) = 12) \approx 0.000015$;
- $P(PL(k,T) = 13) \approx 0.000001$.

### 5.3.3 Observations and Implications

1. Trie Size and Path Length: As the number of addresses in the trie increases, the expected path length also increases. This is consistent with our intuition about the structure of Patricia tries.

2. Concentration of Probabilities: For each trie size, we observe that the probabilities are concentrated around a small range of path lengths. This suggests that most paths in a Patricia trie of random addresses have similar lengths.

3. Logarithmic Growth: The most probable path length grows logarithmically with the number of addresses, which is consistent with the expected $O(\log N)$ performance of Patricia trie operations.

4. Rare Long Paths: Very long paths (close to the maximum of 41) have extremely low probabilities, even in large tries. This indicates that the Patricia trie structure effectively compresses shared prefixes.

5. Implications for Performance: The concentration of path lengths around the expected value suggests that performance of lookup, insertion, and deletion operations in Patricia tries should be relatively consistent across different addresses.

These calculations and observations provide valuable insights into the expected structure of Patricia tries containing random Ethereum addresses. They form a foundation for further analysis of trie performance and can guide optimization strategies for Ethereum's state management system.

## 6. Expected Tree Structure

The analysis of the expected structure of Patricia tries containing randomly generated Ethereum addresses is crucial for understanding the performance characteristics and optimization opportunities in Ethereum's state management system. This section provides a comprehensive examination of the tree structure, including the analysis of path length distributions, and the derivation of average path length.

### 6.1 Theoretical Derivation

The probability mass function (PMF) for path length $k$ in a trie with $N$ addresses, derived in Section 4, is:

$$P(PL = k) = \left(1 - \left(\frac{1}{16}\right)^k \cdot \frac{15}{16}\right)^N - \left(1 - \left(\frac{1}{16}\right)^{k-1} \cdot \frac{15}{16}\right)^N.$$

The cumulative distribution function (CDF) for path length $k$ is:

$$F(k) = P(PL \leq k) = 1 - \left(1 - \left(\frac{1}{16}\right)^k \cdot \frac{15}{16}\right)^N.$$

The average path length is a critical metric for understanding the overall performance of operations on a Patricia trie. The expected path length, $E[PL]$, can be derived using the probability mass function:

$$E[PL] = \sum_{k=2}^{41} k \cdot P(PL = k).$$

Substituting our PMF:

$$E[PL] = \sum_{k=2}^{41} k \cdot \left( \left(1 - \left(\frac{1}{16}\right)^k \cdot \frac{15}{16}\right)^N - \left(1 - \left(\frac{1}{16}\right)^{k-1} \cdot \frac{15}{16}\right)^N \right).$$

### 6.2 Numerical Evaluation

Let's evaluate the average path length for different trie sizes:

1. Small Trie ($n = 100$): $E[PL] \approx 2.328879$;
2. Medium Trie ($n = 10,000$): $E[PL] \approx 4.041428$;
3. Large Trie ($N = 1,000,000$): $E[PL] \approx 5.649078$;
5. Ethereum Tree ($N = 300,000,000$): $E[PL] \approx 7.717012$.

### 6.3 Asymptotic Behavior

As $n$ approaches infinity, the average path length grows logarithmically:

$$\lim_{n \to \infty} \frac{E[PL]}{\log_{16}(N)} = 1.$$

This asymptotic behavior confirms the logarithmic time complexity of operations in Patricia tries.

### 6.4 Implications for Ethereum State Management

- Scalability: The logarithmic growth of average path length suggests good scalability for Ethereum's state trie as the number of addresses increases.
- Consistency: The relatively low variance in path lengths implies consistent performance across different addresses.
- Optimization Opportunities: Understanding the expected structure allows for targeted optimizations, such as caching strategies focused on the most common path lengths.

This analysis of the expected structure of Patricia tries containing random Ethereum addresses provides a solid foundation for understanding and optimizing the performance of Ethereum's state management system. The derived formulas and quantitative estimations offer valuable insights for both theoretical analysis and practical implementation considerations.

# 7. Experimental Validation of Patricia Trie Model for Ethereum Addresses

## 7.1 Introduction

The theoretical model of Patricia tries for Ethereum addresses, as developed in previous sections, provides a foundation for understanding the structure and performance characteristics of Ethereum's state storage system. To validate this model and assess its practical applicability, we conducted a series of comprehensive computational experiments. This section presents the methodology, results, and analysis of these experiments, offering empirical support for our theoretical predictions and insights into the behavior of Patricia tries at various scales.

## 7.2 Experimental Methodology

### 7.2.1 Simulation Environment

All experiments were conducted using a Python-based simulation environment. The source code for the simulation is available in a public Google Colab notebook [20], ensuring reproducibility and facilitating independent verification of our results.

### 7.2.2 Address Generation

We generated random Ethereum addresses using a cryptographically secure pseudo-random number generator (CSPRNG) to simulate private keys, followed by a Keccak-256 hash operation to derive the corresponding address. This process ensures that the generated addresses closely mimic the statistical properties of real Ethereum addresses.

### 7.2.3 Trie Construction and Analysis

For each experiment, we constructed a Patricia trie by sequentially inserting the generated addresses. After construction, we performed a depth-first traversal of the trie to calculate the path length for each inserted address.

### 7.2.4 Experimental Parameters

We conducted experiments with four different trie sizes:

1. Small trie: 100 addresses
2. Medium trie: 1,000 addresses
3. Large trie: 10,000 addresses
4. Very large trie: 100,000 addresses

For each trie size, we performed 10 independent simulations to ensure statistical significance and to account for variability in the random address generation process.

### 7.2.5 Statistical Analysis

For each experiment, we computed the following metrics:

1. Path length distribution
2. Average path length
3. Chi-square goodness-of-fit test comparing observed frequencies with theoretical probabilities

## 7.3 Experimental Results

### 7.3.1 Path Length Distributions

Tables 1-4 present the comparison between theoretical and experimental probabilities for different path lengths across various trie sizes.

**Table 1:** Path length distribution for 100 addresses

| Path Length | Theoretical Prob. | Experimental Prob. | Difference |
|---|---|---|---|
| 1 | 0.002386 | 0.000000 | 0.002386 |
| 2 | 0.690504 | 0.684000 | 0.006504 |
| 3 | 0.284479 | 0.306000 | 0.021521 |
| 4 | 0.021201 | 0.010000 | 0.011201 |
| 5 | 0.001340 | 0.000000 | 0.001340 |
| 6 | 0.000084 | 0.000000 | 0.000084 |

**Table 2:** Path length distribution for 1,000 addresses

| Path Length | Theoretical Prob. | Experimental Prob. | Difference |
|---|---|---|---|
| 2 | 0.025506 | 0.020200 | 0.005306 |
| 3 | 0.769895 | 0.763500 | 0.006395 |
| 4 | 0.190395 | 0.214300 | 0.023905 |
| 5 | 0.013310 | 0.002000 | 0.011310 |
| 6 | 0.000838 | 0.000000 | 0.000838 |
| 7 | 0.000052 | 0.000000 | 0.000052 |

**Table 3:** Path length distribution for 10,000 addresses

| Path Length | Theoretical Prob. | Experimental Prob. | Difference |
|---|---|---|---|
| 3 | 0.101360 | 0.085860 | 0.015500 |
| 4 | 0.765349 | 0.786600 | 0.021251 |
| 5 | 0.124390 | 0.126340 | 0.001950 |
| 6 | 0.008342 | 0.001200 | 0.007142 |
| 7 | 0.000524 | 0.000000 | 0.000524 |
| 8 | 0.000033 | 0.000000 | 0.000033 |

**Table 4:** Path length distribution for 100,000 addresses

| Path Length | Theoretical Prob. | Experimental Prob. | Difference |
|---|---|---|---|
| 4 | 0.239184 | 0.217824 | 0.021360 |
| 5 | 0.675289 | 0.712484 | 0.037195 |
| 6 | 0.079954 | 0.069361 | 0.010593 |
| 7 | 0.005223 | 0.000331 | 0.004892 |
| 8 | 0.000327 | 0.000000 | 0.000327 |
| 9 | 0.000020 | 0.000000 | 0.000020 |

### 7.3.2 Average Path Lengths

Table 5 compares the theoretical and experimental average path lengths for each trie size.

**Table 5:** Comparison of theoretical and experimental average path lengths

| Number of Addresses | Theoretical Avg. | Experimental Avg. | Difference |
|---|---|---|---|
| 100 | 2.33 | 2.33 | 0.00 |
| 1,000 | 3.19 | 3.20 | 0.01 |
| 10,000 | 4.04 | 4.04 | 0.00 |
| 100,000 | 4.85 | 4.85 | 0.00 |

### 7.3.3 Statistical Significance

Table 6 presents the results of chi-square goodness-of-fit tests comparing the observed frequencies with the theoretical probabilities.

**Table 6:** Chi-square goodness-of-fit test results

| Number of Addresses | Chi-square Statistic | p-value |
|---|---|---|
| 100 | 0.011423 | 1.0 |
| 1,000 | 0.014662 | 1.0 |
| 10,000 | 0.009664 | 1.0 |
| 100,000 | N/A* | N/A* |

*Note: The chi-square test for 100,000 addresses resulted in a divide-by-zero error due to extremely small expected frequencies for certain path lengths. This limitation is discussed in Section 7.4.3.

# 8 Discussion

## 8.1 Accuracy of Path Length Distributions

The experimental results demonstrate remarkable agreement with the theoretical probabilities across all trie sizes. The differences between theoretical and experimental probabilities are generally small, with the largest discrepancy being approximately 3.7 percentage points (for path length 5 in the 100,000 address case, Table 4).

This close alignment validates our theoretical model's ability to predict the structure of Patricia tries containing random Ethereum addresses. The accuracy persists across different trie sizes, from 100 to 100,000 addresses, indicating the model's robustness and scalability.

## 8.2 Confirmation of Key Theoretical Predictions

- Logarithmic Growth of Path Lengths: The experimental data confirms that the most common path length increases logarithmically with the number of addresses. This is evident from the shift in the mode of the distribution from path length 2 for 100 addresses (Table 1) to path length 5 for 100,000 addresses (Table 4).

- Right-Skewed Distribution: The experimental results consistently show a right-skewed distribution of path lengths, with rapidly decreasing probabilities for longer paths. This characteristic is crucial for understanding the worst-case scenarios in trie operations and for optimizing storage and retrieval mechanisms.

- Precision of Average Path Length Predictions: The experimental average path lengths show remarkable agreement with the theoretical predictions (Table 5). The maximum difference observed is only 0.01, occurring in the 1,000 address case. This precision is particularly noteworthy given the stochastic nature of the address generation process and trie construction.

## 8.3 Statistical Significance and Model Validation

The chi-square goodness-of-fit tests (Table 6) provide strong statistical support for our theoretical model. The high p-values (1.0) for trie sizes up to 10,000 addresses indicate no significant difference between the observed and expected distributions.

It's important to note that for the 100,000 address case, the chi-square test encountered computational limitations due to extremely small expected frequencies for certain path lengths. This highlights a challenge in validating probabilistic models for very large tries, where the occurrence of certain path lengths becomes exceedingly rare.

## 8.4 Implications for Ethereum State Management

1. Scalability: The consistent accuracy of our model across different trie sizes suggests that Ethereum's state trie structure scales predictably. This predictability is crucial for estimating system performance as the state size grows.

2. Optimization Opportunities: The validated path length distributions provide insights for optimizing node storage and retrieval. For instance, caching strategies could prioritize nodes at the most common path lengths, potentially improving average access times.

3. Worst-Case Scenarios: The experimental confirmation of the right-skewed nature of path length distributions allows for more accurate estimation of worst-case scenarios in trie operations. This information is valuable for ensuring consistent performance in time-critical blockchain operations.

4. State Growth Modeling: The accuracy of our model in predicting average path lengths can inform more precise estimates of state growth rates and associated storage requirements as the Ethereum network expands.

**8.5 Limitations and Future Work**

While our experimental results strongly support the theoretical model, several areas warrant further investigation:

- Larger Scale Testing: Experiments with even larger tries (millions to billions of addresses) could further validate the model's scalability. However, such tests would require significant computational resources and may face challenges in statistical validation due to the rarity of certain path lengths.

- Dynamic Scenarios: Our current experiments focus on static tries. Future work could investigate trie behavior under dynamic insert/delete operations [6], [21], providing insights into real-world performance in actively used blockchain systems.

- Non-Uniform Distributions: The current model assumes uniformly distributed random addresses. Analysis of tries constructed from non-uniformly distributed addresses (e.g., contract-generated addresses or addresses with specific patterns) could extend the model's applicability to more diverse scenarios.

- Hardware-Specific Performance: While our model accurately predicts structural properties of Patricia tries, empirical studies of trie operations on various hardware configurations could bridge the gap between theoretical predictions and practical performance metrics.

- Advanced Statistical Techniques: For very large tries, where traditional goodness-of-fit tests face limitations, developing more sophisticated statistical methods for model validation could enhance the rigor of future analyses.

Our experimental results provide strong empirical support for the theoretical analysis of Patricia tries containing randomly generated Ethereum addresses. The close agreement between predicted and observed path length distributions, coupled with the precision in average path length predictions, validates the model's accuracy and utility.

This validated model offers a solid foundation for further research and optimization in blockchain state management systems, particularly in the context of Ethereum and similar platforms. It provides a powerful tool for predicting trie structure and performance characteristics, informing design decisions in Ethereum clients, and guiding optimization efforts in areas such as state caching, pruning strategies, and node layout in storage systems.

The robustness of the model across different trie sizes demonstrates its value in understanding and optimizing Ethereum's state management at various scales, from small test networks to large-scale production environments. As Ethereum continues to grow and evolve, this theoretically grounded and experimentally validated understanding of its fundamental data structures will be crucial for ensuring the scalability and efficiency of the network.

# 9. Conclusion

## 9.1 Summary of Key Findings

This study has provided a comprehensive theoretical and empirical analysis of Patricia tries as used in Ethereum's state management system. Our key findings include:

- Probabilistic Model Validation: We developed and experimentally validated a probabilistic model for path length distributions in Patricia tries containing random Ethereum addresses. The model accurately predicts trie structures across various scales, from 100 to 100,000 addresses.

- Logarithmic Scaling: Both theoretical analysis and experimental results confirm the logarithmic scaling of average path lengths with respect to the number of addresses. This property is crucial for Ethereum's scalability, ensuring efficient state management as the network grows.

- Path Length Distribution Characteristics: We identified and verified the right-skewed nature of path length distributions, providing insights into worst-case scenarios and informing optimization strategies.

- Precision in Predictions: Our model demonstrated high precision in predicting average path lengths, with discrepancies between theoretical and experimental results not exceeding 0.01 across all tested scales.

- Statistical Robustness: Chi-square goodness-of-fit tests provided strong statistical support for our model, with p-values of 1.0 for trie sizes up to 10,000 addresses, indicating no significant difference between observed and expected distributions.

- Structural Insights: The analysis revealed the concentration of nodes at specific levels of the trie, offering valuable insights for optimizing storage and retrieval mechanisms.

These findings contribute to a deeper understanding of Ethereum's fundamental data structures and provide a solid foundation for future optimizations and scalability improvements.

## 9.2 Potential Directions for Future Research

While our study provides significant insights, several avenues for future research emerge:

- Extreme Scale Analysis: Investigating Patricia trie behavior for extremely large address spaces (billions of addresses) could provide insights into Ethereum's long-term scalability prospects.

- Dynamic Trie Behavior: Analyzing trie performance under frequent insertions and deletions could offer insights into real-world blockchain dynamics and inform adaptive optimization strategies.

- Non-Uniform Address Distributions: Extending the model to account for non-uniform address distributions, such as those resulting from smart contract interactions or specific address generation patterns, could enhance its practical applicability.

- Hardware-Specific Optimizations: Conducting empirical studies on various hardware configurations could bridge the gap between theoretical models and practical implementation, potentially leading to hardware-specific optimization strategies.

- Comparative Analysis: Comparing Patricia tries with alternative data structures for state management could identify potential improvements or hybrid approaches for blockchain systems.

- Pruning and State Expiry Strategies: Leveraging the structural insights from our model to develop efficient state pruning and expiry mechanisms could address Ethereum's state growth challenges.

- Machine Learning Integration: Exploring the potential of machine learning techniques for predicting trie structure evolution and optimizing state management based on historical data and network patterns.

- Cross-Chain Applicability: Investigating the applicability of our model and findings to other blockchain systems that use similar state management structures could contribute to broader blockchain scalability solutions.

These directions for future research hold the potential to further enhance the efficiency, scalability, and adaptability of Ethereum and similar blockchain systems, contributing to the ongoing evolution of decentralized technologies.